\documentclass[usenatbib,graphicx,referee]{mn2e}
\usepackage{amsmath,amssymb}
   \usepackage[pdftex]{graphicx}

\def\lta{\lower2pt\hbox{$\buildrel {\scriptstyle <}
   \over {\scriptstyle\sim}$}}
\def\gta{\lower2pt\hbox{$\buildrel {\scriptstyle >}
   \over {\scriptstyle\sim}$}}
\def\G{\Gamma}
\def\g{\gamma}

\def\Fopt{F_{opt}}
\def\nuopt{\nu_{opt}}
\title[]{Observational Limits on Inverse Compton Processes in GRBs }
\author[]{Tsvi Piran$^{1}$\thanks{tsvi@phys.huji.ac.il},
Re'em Sari$^{1,2}$\thanks{sari@tapir.caltech.edu}, Yuan-Chuan
Zou$^{1}$\thanks{zou@phys.huji.ac.il}\\
\noindent \normalsize{$^{1}$Racah Institute for Physics,
The Hebrew University, Jerusalem, 91904, Israel}\\
\noindent \normalsize{$^{2}$Theoretical Astrophysics, Caltech,
Pasadena CA 91125, USA}}
\begin{document}\maketitle

\begin{abstract}
Inverse Compton (IC) scattering is one of  two viable mechanisms
that can produce the prompt non-thermal soft gamma-ray emission in
Gamma-Ray Bursts. IC requires low energy seed photons and a
population of relativistic electrons that upscatter them. The same
electrons will  upscatter the gamma-ray photons to even higher
energies in the TeV range. Using the current upper limits on the
prompt optical emission we show  that under general conservative
assumption the IC mechanism suffers from an ``energy crisis".
Namely, IC will over-produce a very high energy component that would
carry much more energy than the observed prompt gamma-rays, or
alternatively it will require a low energy seed that is more
energetic  than the prompt $\gamma$-rays. Our analysis is general
and it makes no assumptions on the specific mechanism that produces
the relativistic electrons population.
\end{abstract}

\bigskip\medskip

\section{Introduction}
The mechanism that produces the prompt gamma-ray emission in Gamma
Ray Burst (GRBs) is still uncertain. The non-thermal character of
the spectrum points out towards Inverse Compton (IC) and Synchrotron
as the two natural candidates. The latter become, somehow, the
``standard" process but the former remained always a serious
alternative \citep[][and
others]{s94,sd95,snp96,sp97,w97,glcr00,sp04,k07}. The observations
of numerous bursts with low energy spectral slopes that are
inconsistent with synchrotron  \citep{c97,p98,glcr00,p02} provided
additional motivation to consider IC.  Recently  \cite{km08} have
shown further inconsistency with the overall synchrotron model and
suggested that  Synchrotron Self-Compton (SSC) can resolve some of
these problems.

The recent observations of a naked eye optical flash from GRB080319b
\citep{r08,b08,d08} that coincided in time with the prompt
$\gamma-$ray emission provided further motivation to consider IC as
the source of the prompt $\gamma$-rays. Among the different models
that appeared so far \citep{kp08,fp08,zps08,ywd08}, several favor
models in which the prompt $\gamma$-ray emission is IC of the optical flash
and there have been suggestions that this is generic to many GRBs.

Motivated by these ideas we examine, here,  the possibility that IC
is the source of the prompt soft $\gamma$-ray emission in GRBs. This
requires  a soft component at the IR-UV range that serves as the
seed for the IC process. The flux of these seed photons is
constrained by observations (or upper limits) of the prompt optical
emission. GRB 990123 \citep{ab99} and GRB 080319B \citep{r08} are
rare exceptions with very strong optical emission,  $\sim 9$ and
$\sim 5.3$ mag respectively. However most bursts are much dimer
optically with observations or upper limits around 14 mag
\citep{ya07}. This should be compared with fluxes of $mJy$ in soft
gamma rays  for a modest burst.  What is important, in this work is
the flux ratio $F_\g /F_{opt}$ which is typically larger than 0.1
during the peak soft gamma emission \citep{ya07}.

The basic problem of the IC  model can be explained simply. If the
low energy seed  emission is in the optical, while the observed soft
$\gamma$-ray spectrum is the first IC component, then second IC
scatterings would create a TeV component. Upper limits or
observations of the prompt optical signal show that the $Y$
parameter, i.e. the ratio between the energy in the first IC
component to that in the low energy seed photons is very large,
typically greater than thousands. Theory would then show that the
second IC component in the TeV range would carry an even larger
amount of energy, again by a factor of $Y \gg 1$, producing an
``energy crisis" for this model, and possibly violating upper limits
from EGRET (Energetic Gamma-Ray Experiment Telescope)
\footnote{Deeper upper limits on a wider energy range, may
soon come up from Fermi, making our argument stronger.}
\citep{gs05,ans08}. This problem is generic and it does not depend
on the specific details of the overall model.

The above analysis is oversimplified and two factors may alleviate
the energy catastrophe. First, the frequency of the  seed photons
may differ from that where upper limits exist, allowing larger seed
flux and reducing the lower limits on $Y$. Second, the Klein-Nishina
(KN) suppression, which does not affect the first scattering, may
affect the second, resulting in a lower $Y$ parameter for the second
scattering than the first one. In this article, we explore the
parameter space to see weather there exist a regime where a
combination of these two factors allows for less energy in the
second IC component (Typically in the TeV range) than in the
$\gamma$-rays. We find that possible solutions are limited to a very
small region in the parameters space in which the seed photons are
in the IR, the bulk Lorentz factor is very low $(\le 200$) and the
electrons' Lorentz factor is very large ($\ge 2000)$. However, this
solution implies a healthy emission in the IR, while self absorption
limits it. Therefore, when taking self-absorption into account, this
solution is  ruled out as well. A second possible solution exists if
the seed photons are in the UV. This solution requires a very low
electrons' Lorentz factor $\le 100$, and a seed photon flux that
carries comparable energy to the observed prompt $\gamma$-rays.
Furthermore, prompt X-ray observations limit the high energy tail of
the UV component and practically rule out this model.

\begin{figure}
%\vspace{-3cm}
\centerline{\includegraphics[width=0.9\textwidth]{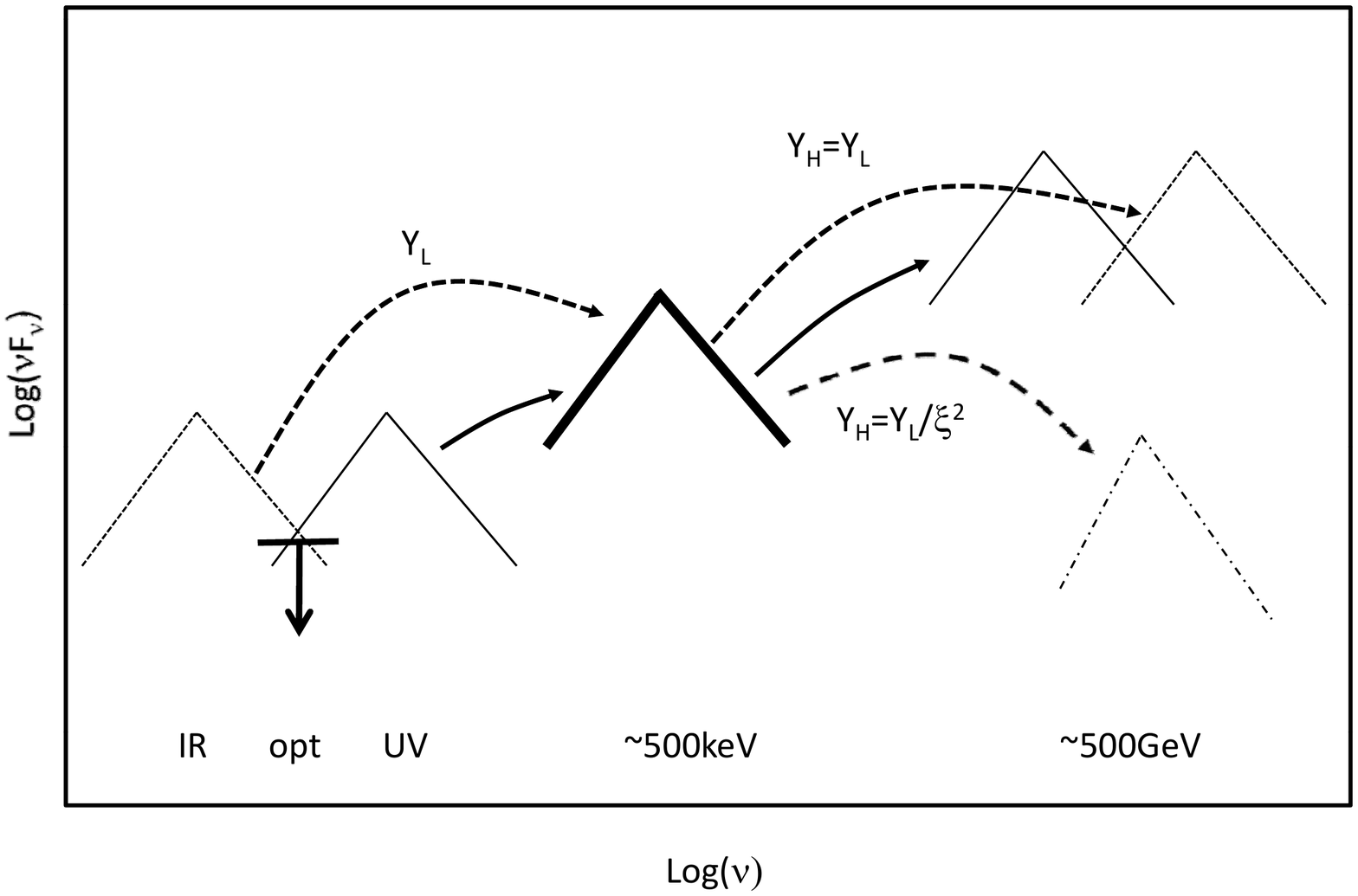}}
\caption{A schematic description of the IC process. Low energy
photons at the IR (marked in dotted lines), optical or UV (marked in
solid thin lines) are IC scattered to produce the observed soft
gamma ray emission (marked in bold lines). A second IC scattering
brings the soft gamma photons to the TeV region. If the initial seed
photons are softer the higher energy component is harder. If the
initial seed is in the IR then the second IC process might be in the
KN regime, in which case this component is suppressed (dashed-dotted
line). The seed low energy emission is constraint by upper limits on
the optical prompt observations (bold solid arrow).}
\label{fig:schematic}
\end{figure}

We take the Lorentz factor of the electrons and the bulk Lorentz
factor as free parameters and we estimate what is the second IC fluence (at TeV
or multi GeV) given the observed prompt gamma-ray flux
and the limits on the prompt optical band.  Most of our analysis is
insensitive to the size of the source, which appears only in the
final section when we estimate the self absorption flux. In our
numerical examples we use very conservative parameters. For example
we use R magnitude of 11.2 as an upper limit on the optical flux,
while many limits are much stronger and the $\gamma$-ray flux we
take,  $10^{-26} {\rm erg\,cm^{-2}\,s^{-1}\,Hz^{-1}}$, is quite
modest. Similarly we use conservative rather rather than
``canonical" values for the spectral slopes.

\section{Basic Equations}

 Consider electrons that move
with a bulk Lorentz factor $\G \gg 1$ while in the bulk (or fluid)
rest frame they have a typical Lorentz factor $\gamma_e \gg 1$ in a
random direction.  We examine IC scattering of seed photons with a
peak frequency $\nu_{\rm seed}$ and a peak flux $F_{\rm seed}$ (both
measured at the observer's rest frame). We assume that the seed
photons are roughly isotropic in the fluid's frame. This would be the
case if the seed photons are  produced by synchrotron radiation in
the bulk, or any other mechanism local to the moving fluid.  We will
consider External IC, in which the seed photons are produced by an
external source elsewhere. For simplicity we assume that all the
photons have the same energy and all the electrons have the same
Lorentz factor. The energy and flux of the scattered photons are:
\begin{equation}
\nu_{IC} = \nu_{\rm seed} \gamma_e^2 \min(1,\xi^{-1}) \label{gamma}
\end{equation}
and
\begin{equation}
 \nu_{IC} F_{IC} = \nu_{seed} F_{\rm seed} ~ Y~  \min(1,\xi^{-2})
\label{flux}
\end{equation}
where $Y\equiv \tau \gamma_e^2$ and $\tau$ are the  Compton
parameter and the optical depth in the Thomson scattering regime.
Note that the unknown optical depth,  $\tau$, is introduced here in
the definition of $Y$ but it is not used elsewhere in the paper. Our
analysis is independent of this unknown factor.  The factor, $\xi$
corresponds to the correction that arises if the scattering is in
the KN region:
\begin{equation}
\xi\equiv {(\gamma_e/\G) h  \nu_{\rm seed}   \over  m_e c^2} > 1.
\label{fKN}
\end{equation}
The expression given in Eq. \ref{fKN} is approximate. Again  this
approximation is sufficient for our purpose.

We consider now the possibility that the prompt gamma-rays arise due
to IC scattering  of a lower energy component.
We use now the observed gamma-ray
flux, $F_\gamma$, and its peak energy, $\nu_\gamma$  and the upper
limits (or detections) of prompt optical emission, $\Fopt$ at
$\nuopt$ to set limits on the IC process.

The peak flux of the low energy component, $F_L$, is at $\nu_L$
which is not necessarily at the observed frequency $\nuopt$. Given
an upper limit on the prompt optical flux, $F_{opt}$ at $\nu_{opt}$
(or on the flux at any other frequency), we can set a limit on $F_L$
if the optical frequency is in the same spectral region as $\nu_L$,
the peak frequency of the lower spectral component of slope $\alpha$:
\begin{equation}
F_L \le(\nu_L/\nu_{opt})^\alpha F_{opt} . \label{limit}
\end{equation}
The equality here and elsewhere holds when $F_{opt}$ corresponds to
a detection and an inequality corresponds to an upper limit. There
are two possibilities, either $\nu_L>\nu_{opt}$ which we call the
``UV solution" or $\nu_L<\nu_{opt}$ which we call the ``IR
solution". Since by definition, the seed photon energy peaks at
$\nu_L$, we must have $\alpha>-1$ in the UV solution and $\alpha<-1$
in the IR solution. Moreover, since the spectrum around $\nu_L$ is
up-scattered to create the familiar Band spectrum \citep{b93} around
$\nu_\gamma$, we can expect $\alpha\approx -1.25$ for the IR
solution and $\alpha \approx 0$ for the UV solution.

As the first IC scattering results in soft $\gamma$-rays, it is
clearly away from the KN regime and we obtain, using Eqs.
(\ref{gamma},\ref{flux},\ref{limit}) a limit on the Compton
parameter $Y_L$, in the first Compton scattering:
\begin{equation}
Y_L \ge {\over }  \left({\nu_\gamma F_\gamma  \over \nu_{opt}
F_{opt}}\right) \left({\nu_{L} \over \nu_{opt}}\right)^{-(1+\alpha)}
. \label{Y}
\end{equation}

Using this limit we turn now to the second order IC component.  This
process will produce photons in the GeV-TeV range. As the scattered
photon is energetic, it might be in the KN regime and we have:
\begin{equation}
\nu_H = \nu_\gamma \left( {\nu_\g \over \nuopt}\right) \left( {\nuopt
\over \nu_L} \right) \min(1,\xi^{-1}) \label{nuH}
\end{equation}
and
\begin{equation}
Y_H \ge \left({\nu_\gamma F_\gamma  \over \nu_{opt} F_{opt}}\right)
\left({\nu_{L} \over \nu_{opt}}\right)^{-(1+\alpha)}
\min(1,\xi^{-2}). \label{YH1}
\end{equation}
$Y_H$ is the ratio of energy emitted in the high energy (TeV) band
and in lower energy gamma-rays (see Fig. 1).

As a conservative numerical example we will use the following
typical parameters: $ F_\gamma=10^{-26}{\rm ergs\,cm^{-2}\,s^{-1}\,Hz^{-1}}$,
$F_{opt}\le 10^{-24}{\rm ergs\, cm^{-2}\,s^{-1}\,Hz^{-1}}$, leading to a ratio of
$F_\g /F_{opt}\ge 0.01$. This optical flux corresponds to R
magnitude 11.2, which is a very conservative upper limit to the
prompt optical emission of most GRBs while the prompt gamma-ray flux
is moderate. We use $\nu_{opt}=8 \cdot 10^{14}$Hz and $h
\nu_{\gamma}= 500 {\rm keV}$ [both energies 
are larger by a factor of $(1+z) \approx 2$ than
the observed frequencies, R band and 250keV]. Thus
$\nu_\g F_\g /(\nuopt F_{opt}) \ge  1500$.
We will use $ \Gamma=300$ and $\g_e
\equiv (\nu_\g/\nu_{opt})^{1/2} \simeq 400 $ for the canonical
values of $\nu_\g$ and $\nuopt$. We find:
\begin{equation}
h \nu_H = 0.08 {\rm TeV} \left({h \nu_\gamma \over 500 {\rm keV} }\right
)\left({\g_e \over 400}\right)^2 \min\left[ 1,{  \G m_e c^2
\over \gamma_e h \nu_\g}\right]
\end{equation}
and
\begin{equation}
 Y_H \ge 1500 \left({ F_\gamma \over 10^{-26}}
  {10^{-24} \over F_{opt} }\right )
 \left({ h \nu_\g \over 500 {\rm\, keV}}
 {8 \cdot 10^{14} {\rm\, Hz} \over \nuopt }\right )
 \left(\nu_L \over \nuopt\right)^{-(1+\alpha)} \min\left[ 1,\left( \G m_e c^2 \over
  \gamma_e h \nu_\g \right)^2
 \right]
\label{EH}
\end{equation}

The essence of the IC problem is the very large value of $Y_H$,
which arises from the fact that the energy released in prompt
gamma-rays is at least a factor of 1500 larger than the energy
released in prompt optical emission (see Eq. \ref{Y}). The large
values of $Y_H$  implies that the energy emitted in the TeV range
exceeds the observed    soft $\gamma$-rays by several orders of
magnitude.

\begin{figure}
%\vspace{-3cm}
\centerline{\includegraphics[width=\textwidth]{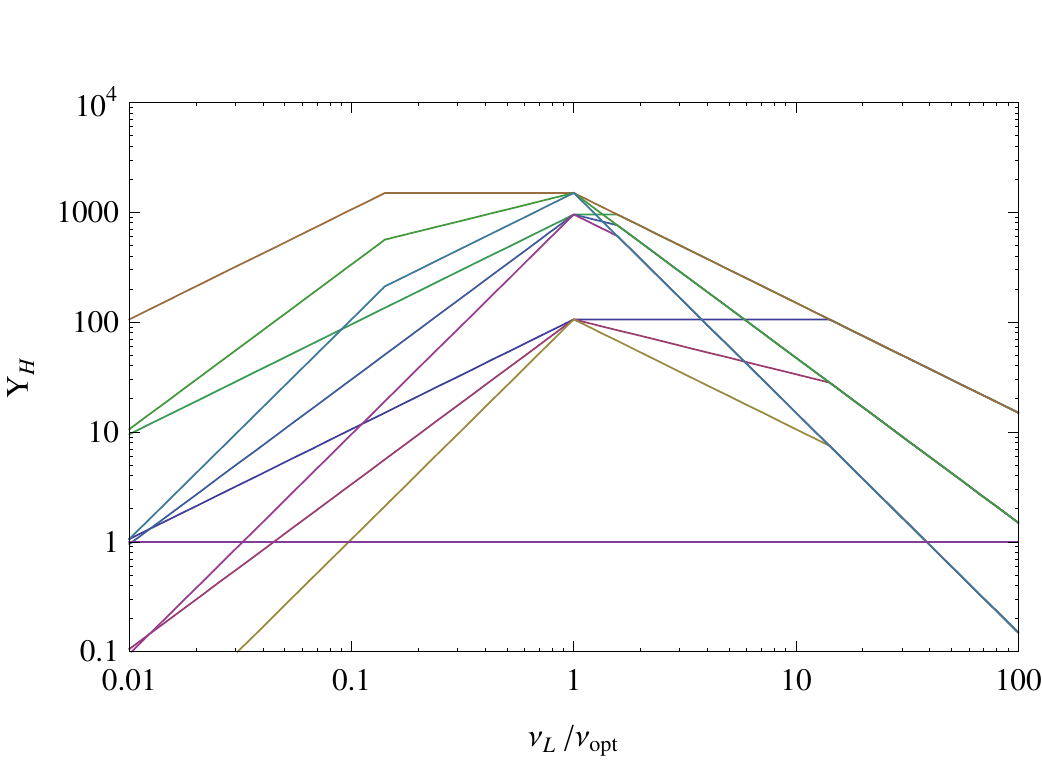}}
\caption{$Y_H$ as a function of $\nu_L/\nuopt$ for $\G =
1000,300,100$ (from top to bottom) and  $\alpha = 0, 0.5, 1$ (from
top to bottom) for $\nu_L>\nuopt$ and $\alpha = -1, -1.5, -2$ for
$\nu_L<\nuopt$. Parameters used in this figure are:
$F_{\gamma}/F_{opt}=0.01$, $\nu_{opt}=8 \cdot 10^{14}$Hz and $h
\nu_{\gamma}=500 {\rm keV}$. The breaks in the lines appear at
$\nu_L=\nuopt$ when we change from negative to positive $\alpha$ and
at the frequency, that depends on $\G$, where the KN correction
begins. \label{fig:Y}}
\end{figure}

Fig. \ref{fig:Y} depicts $Y_H$ as a function of $\nu_L$
 for different values of $\G$ and for different
spectral indices. $Y_H$ peaks when $ \nu_L =\nuopt$. This is
expected as in this case the observed limits on the lower energy
flux are strongest.  If $\nu_L$ increases or decreases  more energy
can be ``hidden" in the lower energy component and the corresponding
$Y_L$ and $Y_H$ will be smaller. Because of a similar reason $Y_H$
decreases when $|\alpha+1|$ increases.

We find two possible regimes for IC solutions that are not over 
producing a high energy (TeV) component. 
The UV solution requires   $\nu_L > 10 \nuopt$ and $\alpha
\ge 1$.  The electrons' Lorentz factor in the UV  solution satisfies
$\gamma_e<100$. The second Compton scattering is not in the KN
regime since $\G>100$ and correspondingly $\xi$ is small.   Since KN
suppression is negligible $Y_L \approx Y_H$ and the total energy,
given by $(1/Y_L + 1 + Y_H) E_\g$, is at least $3 E_\g$. 
UV solutions with $Y_L=Y_H<1$ are therefore also wasteful
as they require a large ($E_\gamma/Y_L$) low energy component.
A second
problem arises, for this solution, with the spectral shape. The
observed low energy spectral index (in the X-ray band) is typically
close to zero, while this solution requires a steeply rising flux
from $\nuopt$ to $\nu_L$. Note that in Fig. \ref{fig:Y} we show
conservatively curves for $\alpha=0, 0.5, 1$ even though the
"canonical" value is 0. Moreover, unless there is a pair loading
(that is if there is one electron per proton), then the low $\gamma_e$ required for the UV
solution implies that the protons carry significantly more energy than
the electrons by at least a factor of $m_p/\gamma_e m_e$. Thus this solution
is a very inefficient.  

\begin{figure}
\centerline{\includegraphics[width=15.cm]{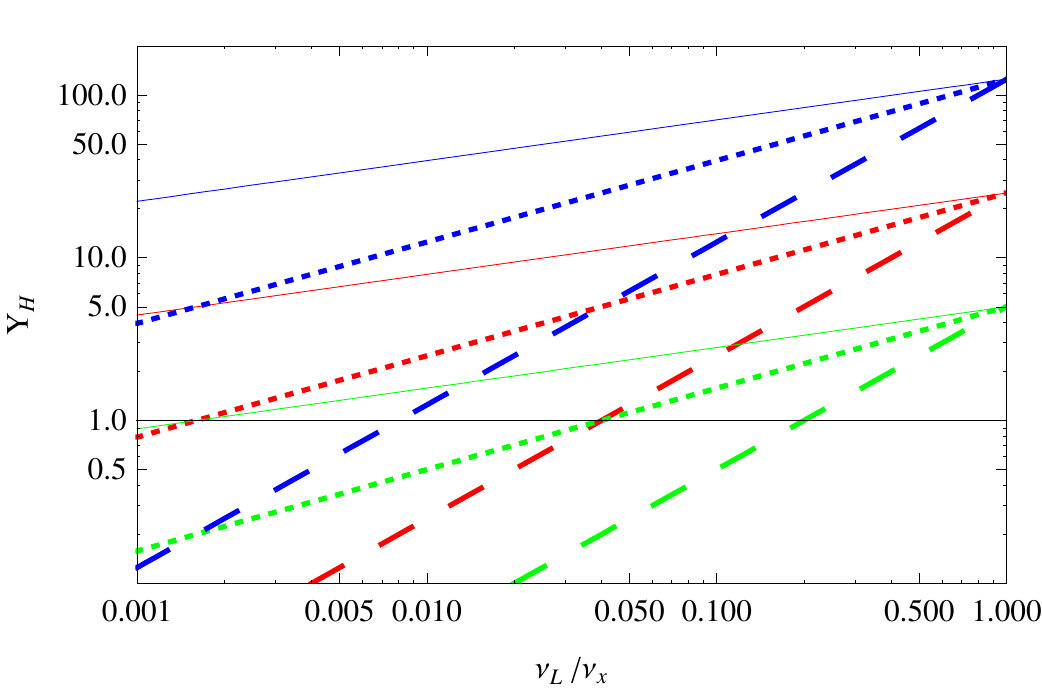}} \caption{$Y_H$ as
a function of $\nu_L/\nu_x$ for $\alpha_1 = -0.5, 0, 0.5$  (blue,
red and green) and $\alpha_2 = -1.25, -1.5, -2$ (solid, dotted and
dashed) for  $h \nu_{x}=20$keV and $h \nu_{\gamma}=500 {\rm keV}$.
The corresponding $\g_e$ range is from 158 at $\nu_L=0.001\nu_x$ to
5 at $\nu_L=\nu_x$. \label{fig:UV}}
\end{figure}

The analysis above is based on the optical limits but for the modest
values of  $\g_e$ needed for the UV solution,  $\nu_L$, the peak
flux frequency of the seed photons becomes large (Eq. \ref{gamma})
and $F_L$ is now limited by prompt soft X-ray observations in
additional to the optical limits. For the discussion below, we use
$\alpha_1$ and $\alpha_2$ as the low energy and high energy spectral
indices, respectively. As stated before, the canonical values are
$\alpha_1=0$ and $\alpha_2=-1.25$ \citep{b93}\footnote{Since we
consider flux rather than photon counts the indices are shifted by 1
relative to \citet{b93}.}. One can estimate the X-ray flux at $\nu_x =
20$ keV directly from the observations at this energy or using the
flux at $\nu_\gamma \approx 500$\,keV and the low energy spectral
slope $\alpha_1$. Recalling that the IC does not change the spectral
slope, we use the same indices both around $\nu_\gamma$ and around
$\nu_L$. Threfore:
\begin{equation}
F_L < ( { \nu_L /\nu_x} )^{\alpha_2} ( { \nu_x /\nu_\g} )^{\alpha_1}
F_\g .
\end{equation}
Using Eq. \ref{flux} we obtain:
\begin{equation}
Y > { \nu_\g^{{\alpha_1}+1}  \nu_x^{{\alpha_2}-{\alpha_1}} \over
\nu_L^{{\alpha_2}+1} } = (\nu_\g/\nu_x)^{{\alpha_1}-{\alpha_2}}
\g_e^{2 ({\alpha_2} +1)} .
\end{equation}
Since the UV solution is not in the KN regime we have $Y=Y_L= Y_H$.
If we take the typical spectral indices below and  above
$\nu_\gamma$ to be $\alpha_1=0$ and $\alpha_2=-1.25$ respectively
\citep{b93}, and we impose the condition $Y \cong 1$ (where the
total energy required is minimized to $3E_\gamma$), we find that
$\g_e> 3000$ or $\nu_L <\nuopt$ - thus the whole UV regime is ruled
out. This condition depends strongly on the spectral indices:
$\alpha_1$ and $\alpha_2$. Clearly if $\alpha_2$ is smaller (a
steeper drop on the high energy side) $\nu_L$ can be larger and Y is
smaller\footnote{It is intersting to note that $|\alpha_2|$ is large
for GRB 080319b, which might be an IC burst with a UV solution.}. The
limits are depicted in Fig. \ref{fig:UV} for several values of the
spectral indices. One can see that the available X-ray data rules
out the UV solution for most of the phase space.

The IR solution holds for $\nu_L < 0.1 \nuopt =8\cdot 10^{13}$Hz and
$\alpha \le-1.5$. It requires a large electron's Lorentz factor
$\gamma_e \ge 1000$ and a relatively low bulk Lorentz factor
$\Gamma<300$. The solution is deep in the KN regime and the KN
suppression is very significant. It allows for a large amplification
between the IR and the soft $\gamma$-rays and no amplification
between the low energy gamma and the TeV emission. A solution is
possible in a small region of the parameter space if the
high energy spectrum is steep ($\alpha \le -1.5$) - this increases the allowed
flux at $\nu_L$. Such a spectrum above the peak frequency, though steeper
than the canonical $\alpha=-1.25$, is not rare in the observations
of prompt $\gamma$-ray bursts.

\begin{figure}
%\vspace{-3cm}
\centerline{\includegraphics[width=15.cm]{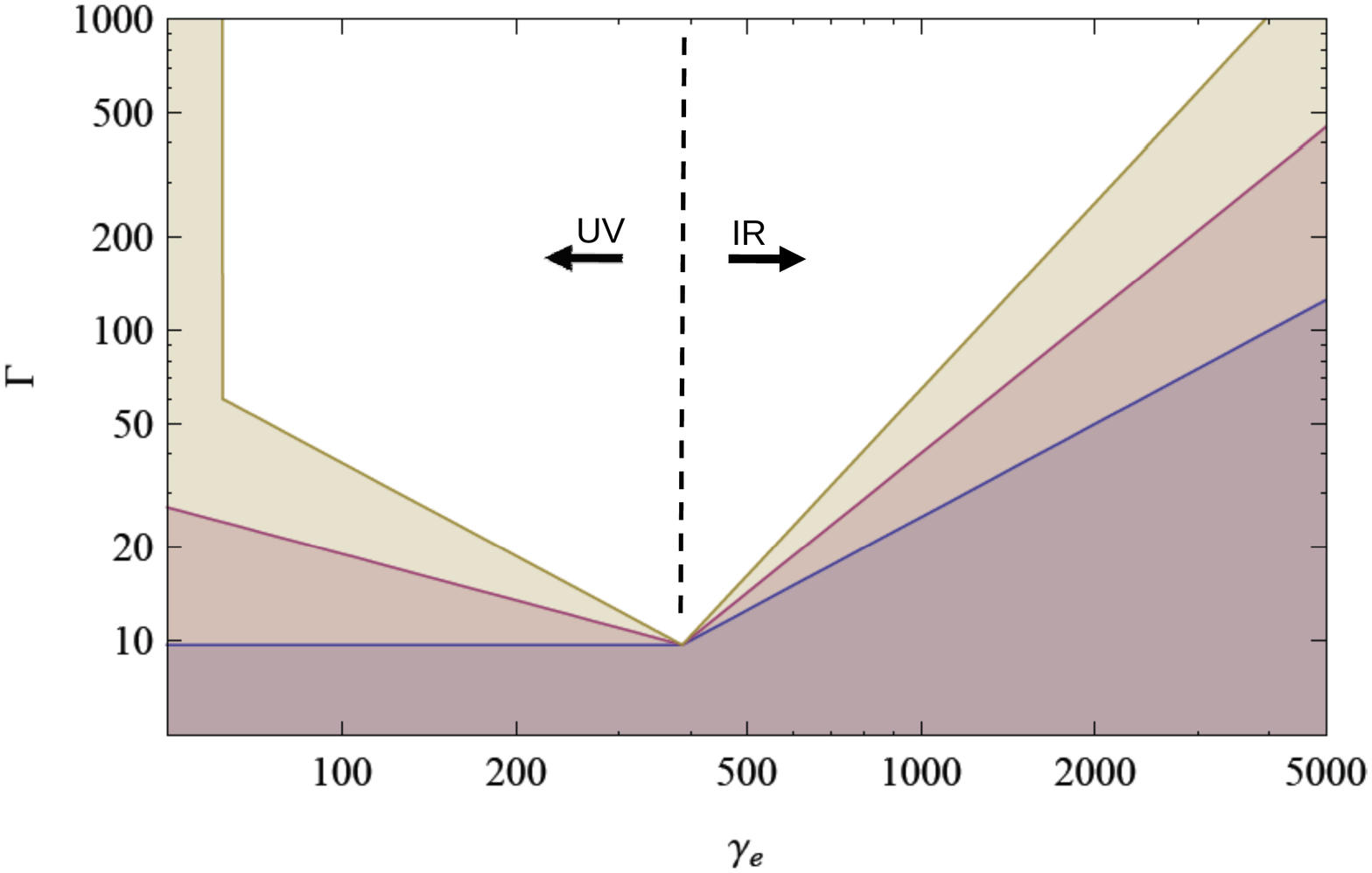}} \caption{The
allowed (colored) phase space in which $Y_H \le 1$. For three
spectral indexes $\alpha = 0, 0.5, 1$ (from bottom to top) for
$\nu_L>\nuopt$ and $\alpha = -1, -1.5, -2$ for $\nu_L<\nuopt$(from
bottom to top). Parameters used are: $F_{\gamma}/F_{opt}=0.01$,
$\nu_{opt}=8 \cdot 10^{14}$Hz and $h \nu_{\gamma}=500 {\rm keV}$.
The $\g_e$ axis corresponds to  values of $\nu_L$ ranging from  $15
\nuopt = 4.8 \cdot 10^{16}$Hz=0.2keV for $\g_e=50$ to $0.006\nuopt=
4.8 \cdot 10^{12}$Hz for $\g_e = 5000$. \label{fig:Gamma}}
\end{figure}

To demonstrate the severity of the constraint we plot
 (Fig. \ref{fig:Gamma}) the ``allowed region" in the ($\g_e,\G$) phase
space for which $Y_H<1$.
%One can solve Eq. \ref{YH1} for a given
%spectral index, to obtain the maximal value of $\G$ so that $Y_H=1$.
%In Fig. \ref{fig:Gamma} we use $Y_H =1$, allowing  equal energy in
%the very high energy emission. If we allow more energy in the very
%high energy regime the limit of $\G$ increase roughly like a square
%root of this factor.
It is remarkable to note that the expected parameter region for
internal shocks $\g_e \approx 500$, $\G \approx 300$ is deep inside
the ruled out region. The parameter expected for external shocks
$\g_e \approx 50,000$ and $\G \approx 300$ are allowed with seed
photon wavelength in the cm range. However, as we show in \S 4, self
absorption limits the amount of energy in such low frequency seed
photons, ruling out this solution. For low values of $\g_e$ the
whole $\G$ range is seemingly allowed.  However, this only happens
at  $\g_e<62,34,10$ for $\alpha = 1.,0.5,0$ respectively, and
therefore conflicts with the soft X-ray observations.

%. This is the UV
%solution for which $Y_L^{-1}\ge 1$ and the seed energy is larger or
%equal than the observed $\nu_\g F_\gamma$.

\section{Pair Avalanche}

In cases when $Y_H>1$ most of the electrons energy is emitted as
very high energy (TeV) gamma-rays. When the scattering is in the KN
regime, that is if Eq. \ref{fKN} holds, the scattered photon, that
has an energy of almost $\gamma_e m_e c^2 $  (in the fluid's rest
frame) and can therefore produce a pair when it encounters a typical
low energy $\gamma$-ray photon with energy $h \nu_\g/\G$ (in this
frame). More specifically, for a head on collision between a photon
with energy $h\nu_\g  = \xi \G m_e c^2 / \g_e $ and an electron with
a Lorentz factor $\g_e$, the energies of the electron and the photon
after the collision are:
\begin{equation}
 h \hat \nu \approx {4 \xi \over 1 + 4 \xi} \G \g_e m_e c^2,
\end{equation}
 and
\begin{equation}
 \hat \g \approx\ {1  \over 1 + 4 \xi} \g_e
\end{equation}
The resulting photon has now enough energy to collide with a photon
with energy $h \nu_\g$ and produce two electrons with a Lorentz
factor:
\begin{equation}
 \check \gamma \approx {4 \xi \over 1 + 4 \xi } {\g_e  \over 2} \approx {\g_e \over 2}
 .
\end{equation}
As the optical depth for pair creation is huge all the scattered
photons will create pairs with typical energy of $\g_e m_e c^2 /2$.
As a result we will have colder electrons and positrons with a ratio
2:1 in higher ($\g_e /2$) and lower ($\g_e/4 \xi$) energies. These
colder electrons and pairs will Inverse Compton scatter more photons
and will produce a second generation of cooler pairs with
$\gamma_e/4$. The process will continue until pair creation will
stop. This will happen when $\tilde \g h \nu_\gamma /\G \approx m_e
c^2$. This situation was considered numerically by
 \citet{c92,sb95,pw05}, and most recently \citet{vp08}.

If the physical conditions, like magnetic field and total number of
particles are fixed $\nu_L$, $\nu_\gamma$ and $\nu_H$ as well as the
corresponding fluxes will vary as a result of the changing electron
energy distribution due to the created pairs. These variations will
be very significant because of the strong dependence (2nd and 4th
powers) of the first two on $\g_e$. The dynamical evolution of such
a system is interesting by itself. However, we are interested, here,
in the final steady state in which $\nu_\gamma$ and $F_\gamma$ are
fixed as the observed quantities. In this case we can search for the
physical parameters that exists in such a steady state. We can
express $\G$ in terms of $\gamma_e$ and $\nu_\gamma$ using the pair
creation threshold criteria (Eq. \ref{fKN}) and we can express
$\nu_L$ in terms of $\nu_\g$ and $\g_e$ (using  Eq. \ref{gamma}).
Given these expressions we can estimate the steady state $Y_H$ as a
function of $\g_e$.

\begin{figure}
%\vspace{-3cm}
\centerline{\includegraphics[width=11.cm]{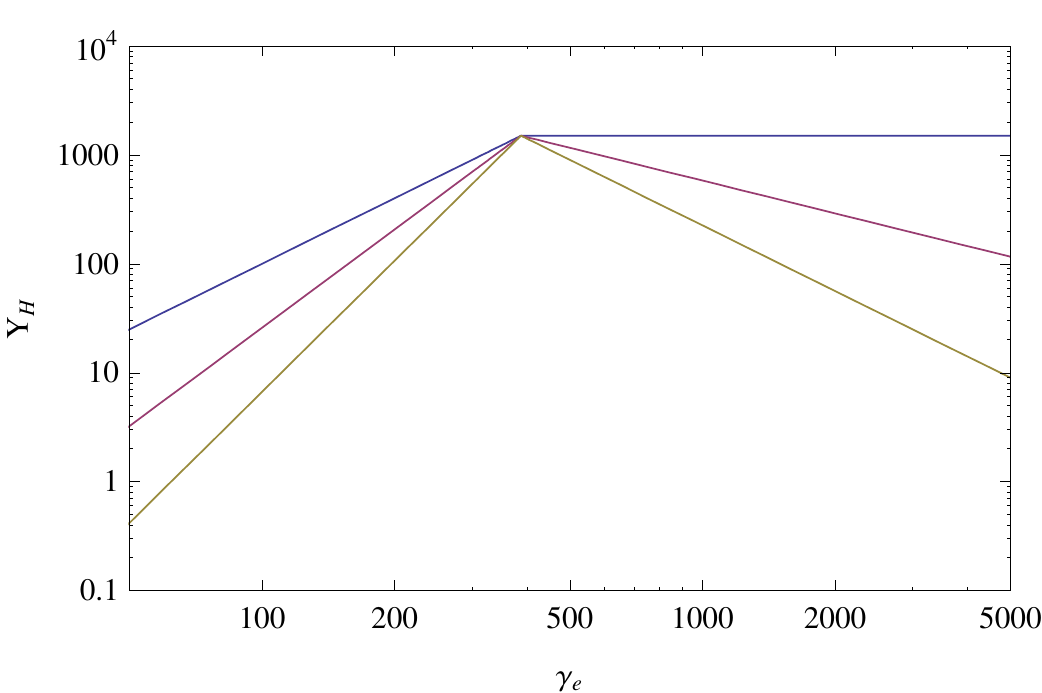}}
\caption{The steady state $Y_H$ as a function of $\g_e$ for a
situation in which pair avalanche leads to  $\gamma_e = m_e c^2
\Gamma / h \nu_\g$. Shown are curves for three different values of
$\alpha = 0,0.5,1$ (from top to bottom) for $\nu_L>\nuopt$ and
$\alpha = -1, -1.5, -2$ for $\nu_L<\nuopt$. The Parameters used are:
$F_{\gamma}/F_{opt}=0.01$, $\nu_{opt}=8 \cdot 10^{14}$Hz and $h
\nu_{\gamma}=500 {\rm keV}$.  The $\g_e$ axis corresponds to  values
of $\nu_L$ ranging from $15 \nuopt = 4.8 \cdot 10^{16}$Hz=0.2keV for
$\g_e=50$ to $0.006\nuopt= 4.8 \cdot 10^{12}$Hz for $\g_e = 5000$.
\label{fig:Pairs}}
\end{figure}

Fig. \ref{fig:Pairs} depicts the resulting $Y_H$ values as a
function of $\g_e$ for different values of $\alpha$. The UV
solution for $\nu_L > 10 \nuopt$ and with rather low values of
$\g_e$ and $\G$ is possible. However this solutions suffers from the
problems discussed earlier. It seems that if we impose the pair
creation threshold conditions the IR solution is ruled out with very
high $Y_H$ values (for any reasonable $\alpha$). However,  as
discussed earlier, there is a region in the parameter space for the
IR solution for which $Y_H \le 1$. In this case only a small
fraction of the energy goes into the high energy photons and it is
possible (depending on time scales) that most of the electrons cool
down rapidly before pair avalanche arises.

\section{The  seed photons and self-absorption}

A natural source of the seed photons is synchrotron emission by the
same electrons that produce the IC emission.
Assuming that this source is indeed synchrotron we can
proceed
%for a given $\nu_L$ or a given $\gamma_e$ and $\alpha$ and
%$\G$ and  use the implied value of $Y$ to estimate the number of
%emitting photons,
and estimate the strength of the magnetic filed and the size of the
emitting region. We can then check if these values are reasonable
within given GRB models. However, we choose a more general approach
and ask whether the large seed flux needed is limited by self
absorption.

\begin{figure}
 \centerline{\includegraphics[width=11.cm]{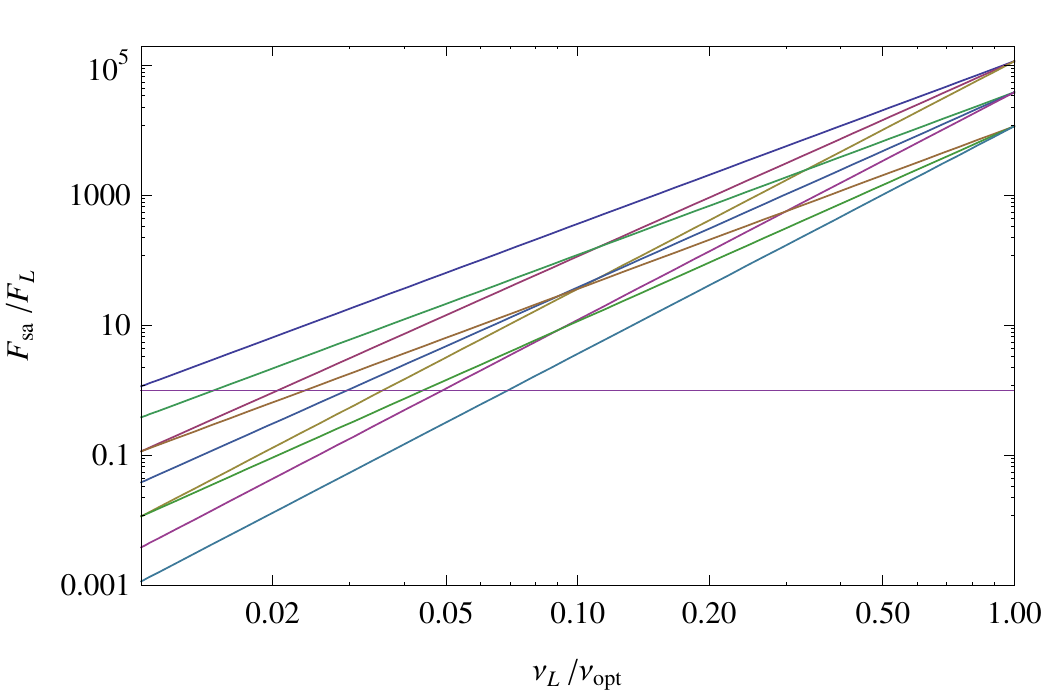}}
\caption{The ratio of the self absorbed flux $F_{sa}$ to the needed
seed flux as function of $\nu_L/\nuopt$ for three values of $\G =
100,300,1000$ (from top to bottom) and three different values of
$\alpha$: $\alpha = -1, -1.5, -2$ (from top to bottom) for
$\nu_L<\nuopt$. Parameters used in this figure are:
$F_{\gamma}/F_{opt}=0.01$, $\nu_{opt}=8 \cdot 10^{14}$Hz and $h
\nu_{\gamma}=500 {\rm keV}$.
 \label{fig:fsa}}
\end{figure}

Self absorption limits the flux at $\nu_L$ to be below the black
body flux, $F_{sa}$, for a local temperature $kT \approx \G \gamma_e
m_e c^2$:
\begin{eqnarray}
%\nonumber
 F_{sa}(\nu_L) & =& {2 \nu_L^2 \over c^2 } { \g_e m_e c^2 } {
R^2 \over 4 \G   d_L^2} \\
%&\approx& 1.3  \cdot 10^{-20} {\rm erg/cm^2/sec/Hz} {(R/10^{17}{\rm
%cm})^2 \over d_L^2(z=1)}{ { (\nu_L}/8 \cdot 10^{14})^{3/2}
%(\nu_\g/500)^{1/2} \over (\G/300) }
%\\
&\approx& \nonumber 1.3  \cdot 10^{-20} {\rm erg\,cm^{-2}\,s^{-1}\,Hz^{-1}}
{(R/10^{17}{\rm cm})^2 \over d_L^2(z=1)} { ( \nu_\g/500)^2 \over
(\g_e/400)^3  (\G/300)} ,
\end{eqnarray}
where $R$ is the radius of the source and $d_L(z=1)$ is the
luminosity distance for z=1. In the following examples we use
conservatively $R=10^{17}$cm as the emission radius of the prompt
emission.

Fig. \ref{fig:fsa} depicts a comparison of this limiting flux,
$F_{sa}$ with the needed flux $F_L =F_\gamma \g_e^2 /Y_L$. For
$\nu_L > 0.1 \nuopt$, $F_{sa}<F_L$. This implies that the electrons
that produce the IC emission cannot produce the lower energy seed
photons. The ratio $F_{sa}/F_L$ decreases with increasing $\G$. It
also decreases when $|\alpha|$ increases. So in most of the region
where $Y_H<1$ (see fig. \ref{fig:Y}) the seed flux is insufficient!

\begin{figure}
\centerline{\includegraphics[width=11.cm]{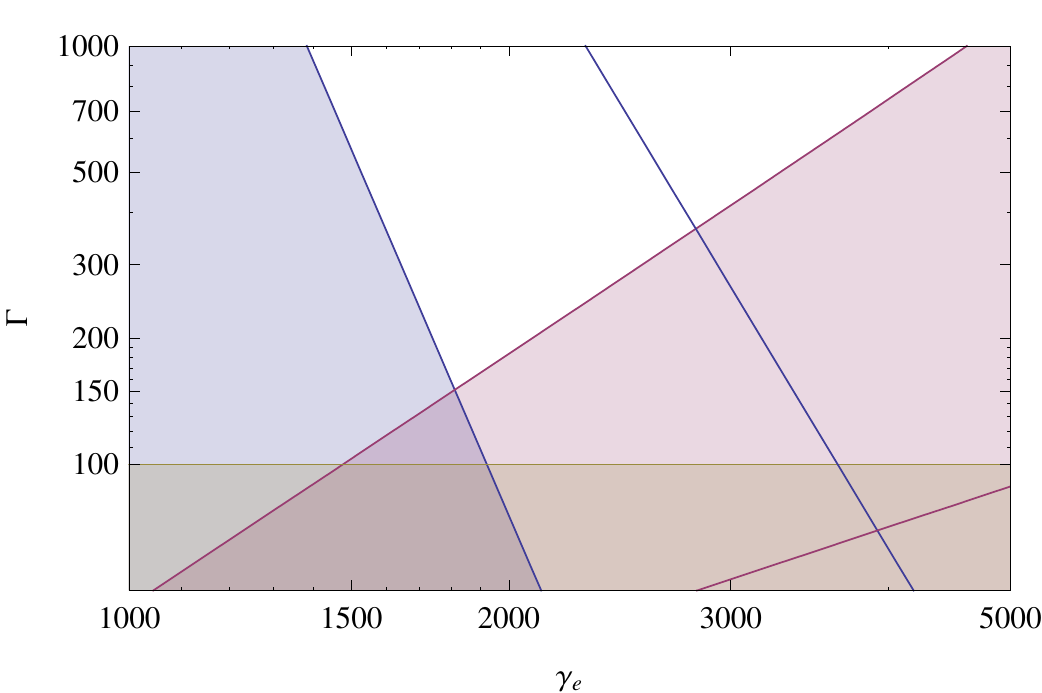}}
\caption{Allowed region for the IR solution in the $(\G,\g_e)$
parameter space. The limit on the left (decreasing curve)
corresponds to the condition $F_{sa}\ge F_L$. The limit on the right
(increasing curve) corresponds to $Y_H=1$.  Also marked is $\G=100$,
which is considered as a minimal value for the bulk Lorentz factor
to resolve the compactness problem \citep{ls01}. The limits are
shown for $\alpha=-2$. (On the right side around $\g_e=4000$ shown
are the corresponding curves for $\alpha =-1$.). The $\g_e$ range
from 1000 to 5000 corresponds to $\nu_L = 1.2 \cdot 10^{14}$Hz to $
\nu_L = 4.8 \cdot 10^{12}$Hz. Parameters used in this figure are:
$F_{\gamma}/F_{opt}=0.01$, $\nu_{opt}=8 \cdot 10^{14}$Hz and $h
\nu_{\gamma}=500 {\rm keV}$. For $\alpha = -2$ an extremely small
region around $\g_e \approx 1800$ (corresponding to $\nu_L = 3.7
\cdot 10^{13}$Hz) and
 $\G \approx 120$
is allowed. \label{fig:fsa2}}
\end{figure}

The combined limits on the $(\G,\g_e)$ parameter space  from self
absorption with $Y_H=1$ are shown in fig. \ref{fig:fsa2}. Only an
extremely small region around $\g_e \approx 1800$ (corresponding to
$\nu_L = 3.7 \cdot 10^{13}$Hz) and $\G \approx 120$ is allowed. This
used a conservative over estimate for the emission radius
$R=10^{17}{\rm\, cm}$. If we use the variability time 
scale $\delta t < 1$sec, with $R \sim \G^2 c \delta t$  and the low values
of $\G$ obtained, $R$ will be much smaller,
invalidating even this
solution. The self absorption limit rules out also the region in the
parameter space that corresponds to external shocks. This solution
requires a very low seed frequency which would have implied a very
small self-absorption limit.

\section{Conclusions}

For a typical GRB, IC has to  amplify the total energy  of a
low energy seed photon flux by a factor of $\approx 1000$ to produce
the observed prompt gamma-ray flux. The same relativistic electrons
will, however, continue and upscatter the gamma-ray flux to very
high energies in the TeV range. In many cases this second generation
IC will be in the Klein-Nishina regime (that is the photon's energy
will be larger than the electrons rest mass, in the electron's rest
frame). This will suppress somewhat the efficiency of conversion of
gamma-rays to very high energy gamma-rays, however it won't stop it
altogether.

Our analysis focused on the case that the low energy seed photons
are produced within the moving region that includes the  IC
scattering relativistic electrons. Such will be the case, for
example, in Synchrotron self-Compton. Related considerations, that
will be published elsewhere, apply when the seed photons are
external and constrain IC processes in this case as well.
The analysis is also limited to the important implicit assumption that the emitting region is 
homogenous. It is possible that very strong inhomogeneities could change this picture.

We have shown that, under quite general conservative assumptions, 
if IC produces the prompt MeV photons, then a second scattering
will over produce a very high (GeV-TeV) prompt component that will
carry significantly more energy than the prompt gamma-rays
themselves. On the  theoretical front such a component will cause an
``energy crisis" for most current progenitor models. From an
observational point  of view, this component is possibly already
ruled out by EGRET upper limits\citep{gs05,ans08}. Fermi should 
very soon put much stronger limit to (or verify) this possibility.
For example, a burst with isotropic energy $E_{\gamma,iso}=10^{53}$erg, 
locating at $z=1$, would produce $\sim 100 Y_H (E_H/10{\rm GeV})$ 
photons detected by Fermi.

One may not over produce a high energy component if the seed photons
are in the UV regime. However, in this case, the needed seed photon
energy should be equal or larger than the observed prompt gamma-ray
energy.  Downwards extrapolation of the X-ray observations put
strong limits on this solution and probably rule it out as well.  Moreover this 
UV solution requires pair loading to be efficient. 

We thank R.  Mochkovitch, E. Nakar, J. Poutanen, P. Kumar and X. F. Wu
for helpful discussions.  The
research was partially supported by the ISF center for excellence in
High Energy Astrophysics, by an Israel-France collaboration grant,
by an IRG grant, by an ATP grant and by the Schwartzman chair.

\end{document}